\def\eg{{e.g., }}

\def\etal{{\it et al.}}

\def\spose#1{\hbox to 0pt{#1\hss}}
\def\lta{\mathrel{\spose{\lower 3pt\hbox{$\mathchar"218$}}
     \raise 2.0pt\hbox{$\mathchar"13C$}}}
\def\gta{\mathrel{\spose{\lower 3pt\hbox{$\mathchar"218$}}
     \raise 2.0pt\hbox{$\mathchar"13E$}}}
\def\ge{\mathrel{\spose{\lower 3pt\hbox{$-$}}
     \raise 2.0pt\hbox{$\mathchar"13E$}}}
\def\le{\mathrel{\spose{\lower 3pt\hbox{$-$}}
     \raise 2.0pt\hbox{$\mathchar"13C$}}}
\def\simgt{\gta}

\documentclass[aps,twocolumn,superscriptaddress]{revtex4}

\usepackage{graphics}

\newcommand {\boom}{{\sc Boomerang}}

\begin{document}

\bibliographystyle{apsrev}

\preprint{}

\title{
First Estimations of Cosmological Parameters from \boom
}

\author{ A.E. Lange}
 \affiliation{California Institute of Technology, Pasadena, CA, USA }
\author{ P.A.R. Ade}
 \affiliation{Queen Mary and Westfield College, London, UK }
\author{ J.J. Bock}
 \affiliation{Jet Propulsion Laboratory, Pasadena, CA, USA }
\author{ J.R. Bond}
 \affiliation{Canadian Institute for Theoretical Astrophysics, University of Toronto, Canada }
\author{ J. Borrill}
 \affiliation{National Energy Research Scientific Computing Center, LBNL, Berkeley, CA, USA }
\author{ A. Boscaleri}
 \affiliation{IROE-CNR, Firenze, Italy }
\author{ K. Coble}
 \affiliation{Dept. of Physics, Univ. of California, Santa Barbara, CA, USA }
\author{ B.P. Crill}
 \affiliation{California Institute of Technology, Pasadena, CA, USA }
\author{ P. de Bernardis}
 \affiliation{Dipartimento di Fisica, Universita' La Sapienza, Roma, Italy}
\author{ P. Farese}
 \affiliation{Dept. of Physics, Univ. of California, Santa Barbara, CA, USA }
\author{ P. Ferreira}
 \affiliation{{Astrophysics, University of Oxford, NAPL, Keble Road, OX2 6HT, UK }}
 \affiliation{Dept. de Physique Theorique, Universite de Geneve, Switzerland}
\author{ K. Ganga}
 \affiliation{California Institute of Technology, Pasadena, CA, USA }
 \affiliation{Physique Corpusculaire et Cosmologie, College de France, 11 place Marcelin Berthelot, 75231 Paris Cedex 05, France }
\author{ M. Giacometti}
 \affiliation{Dipartimento di Fisica, Universita' La Sapienza, Roma, Italy}
\author{ E. Hivon}
 \affiliation{California Institute of Technology, Pasadena, CA, USA }
\author{ V.V. Hristov}
 \affiliation{California Institute of Technology, Pasadena, CA, USA }
\author{ A. Iacoangeli}
 \affiliation{Dipartimento di Fisica, Universita' La Sapienza, Roma, Italy}
\author{ A.H. Jaffe}
 \affiliation{Center for Particle Astrophysics, University of
        California, Berkeley,CA, USA }
\author{ L. Martinis}
 \affiliation{ENEA Centro Ricerche di Frascati, Via E. Fermi 45, 00044
  Frascati, Italy}
\author{ S. Masi}
 \affiliation{Dipartimento di Fisica, Universita' La Sapienza, Roma, Italy}
\author{ P.D. Mauskopf}
 \affiliation{Dept. of Physics and Astronomy, University of Massachussets,
        Amherst, MA, USA }
\author{ A. Melchiorri}
 \affiliation{Dipartimento di Fisica, Universita' La Sapienza, Roma, Italy}
\author{ T. Montroy}
 \affiliation{Dept. of Physics, Univ. of California, Santa Barbara, CA, USA }
\author{ C.B. Netterfield}
 \affiliation{Depts. of Physics and Astronomy, University of Toronto, Canada }
\author{ E. Pascale}
 \affiliation{IROE-CNR, Firenze, Italy }
\author{ F. Piacentini}
 \affiliation{Dipartimento di Fisica, Universita' La Sapienza, Roma, Italy}
\author{ D. Pogosyan}
 \affiliation{Canadian Institute for Theoretical Astrophysics, University of Toronto, Canada }
\author{ S. Prunet}
 \affiliation{Canadian Institute for Theoretical Astrophysics, University of Toronto, Canada }
\author{ S. Rao}
 \affiliation{Istituto Nazionale di Geofisica, Roma,~Italy }
\author{ G. Romeo}
 \affiliation{Istituto Nazionale di Geofisica, Roma,~Italy }
\author{ J.E. Ruhl}
 \affiliation{Dept. of Physics, Univ. of California, Santa Barbara, CA, USA }
\author{ F. Scaramuzzi}
 \affiliation{ENEA Centro Ricerche di Frascati, Via E. Fermi 45, 00044
  Frascati, Italy}
\author{ D. Sforna}
 \affiliation{Dipartimento di Fisica, Universita' La Sapienza, Roma, Italy}


\begin{abstract}
The anisotropy of the cosmic microwave background radiation contains
information about the contents and history of the universe.  We report
new limits on cosmological parameters derived from the angular power
spectrum measured in the first Antarctic flight of the \boom \ experiment.
Within the framework of inflation-motivated adiabatic cold dark matter
models, and using only weakly restrictive prior probabilites on the age of
the universe and the Hubble expansion parameter $h$, we find that the 
curvature is consistent with flat and that the
primordial fluctuation spectrum is consistent with scale invariant, in
agreement with the basic inflation paradigm.  We find that the data
prefer a baryon density $\Omega_b h^2$ above, though similar to, the
estimates from light element abundances and big bang nucleosynthesis.
When combined with large scale structure observations, the \boom \
data provide clear detections of both dark matter and dark energy
contributions to the total energy density $\Omega_{\rm {tot}}$,
independent of data from high redshift supernovae.
\end{abstract}

\pacs{ }

\maketitle



\begin{figure}[t!]
  \resizebox{!}{3.3in}{
  \includegraphics{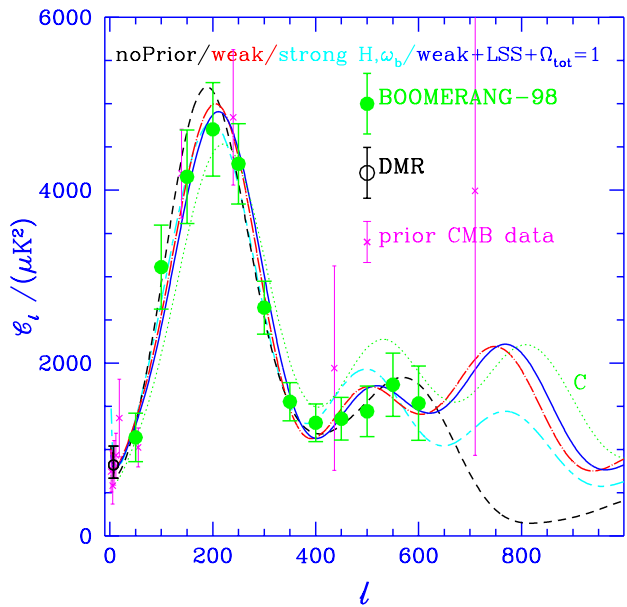}
  }
\caption{\protect\small
  CMB angular power spectra, ${\cal C}_\ell \equiv \ell
(\ell+1)\langle \vert T_{\ell m}\vert^2 \rangle/(2\pi)$, where the
$T_{\ell m}$ are the multipole moments of the CMB temperature.  The
closed (green) circles show the B98 data.  The magenta crosses 
are a
radical compression of all the data prior to B98 into optimal
bandpowers~\protect{\cite{BJK98,toco98,mauskopf99}}, 
showing the qualitative improvement provided
by B98 except in the $\ell \lta 20$ DMR regime, where the
COBE data are represented as a single bandpower (open black circle). 
(Note that the B98 and prior CMB points at $\ell =150$ lie 
on top of each other.) The
smooth curves depict power spectra for several maximum likelihood
models with different priors chosen from
Table~\protect\ref{parameters}, with
($\Omega_{\rm {tot}},\omega_b,\omega_c,\Omega_\Lambda,n_s,\tau_c$) as
follows: P1, short dashed line,(1.3,0.10,0.80,0.6,0.80,0.025); P4,
dot-dashed line,(1.15,0.03,0.17,0.4,0.925,0); P8, short-long dashed
line,(1.05,0.02,0.06,0.90,0.825,0); P11, solid
line,(1.0,0.03,0.17,0.70,0.95,0.025).  These curves are all 
reasonable fits to the B98+COBE data.  For
comparison, we plot a $H_0=68$, $\Omega_\Lambda=0.7$ ``concordance
model'' which does not fit (dotted line labelled C), with parameters
(1.0,0.02,0.12,0.70,1.0,0).}
\label{powerspectrum}  
\end{figure}

The angular power spectrum ${\cal C}_\ell$ of temperature anisotropy
in the cosmic microwave background (CMB) is a powerful probe of the
content and nature of the universe.  The DMR instrument on the COBE
satellite measured ${\cal C}_\ell$ for multipoles $\ell \lta 20$,
corresponding to angular scales $\simgt 7^\circ$\cite{Bennett96}.  
Significant experimental effort by many
groups focusing on smaller angular scales, when
combined~\cite{BJK98,toco98,mauskopf99}, led to the ${\cal C}_\ell$
estimates in the $\ell$ bands marked with x's in
Figure~\ref{powerspectrum}, which indicate a peak 
at $\ell \sim 200$\cite{knox2000}.
It has long been recognized that if ${\cal C}_\ell$ can be determined
with high precision over these angular scales, parameters such as the
total energy density and baryon content of the universe, and the shape
of the primordial power spectrum of density fluctuations, can be
accurately measured~\cite{BET97}.  The most recently published \boom \
angular power spectrum shown in Figure~\ref{powerspectrum} represents
a qualitative step towards such high precision~\cite{debernardis00}
(hereafter, B98).

The data define a strong peak at $\ell \sim 200$. The steep
drop in power from $\ell \sim 200$ to $\ell \sim 400$ is consistent
with the structure expected from acoustic oscillations in adiabatic
cold dark matter (CDM) models of the universe, but is not consistent
with the locations and widths of peaks expected in the simplest cosmic
string, global topological defect, and isocurvature perturbation
models~\cite{defects}. The data at higher $\ell$ show strong detections
which limit the height of a second peak, but are consistent with the height
expected in many CDM models.

In this paper,  we concentrate on determining a set of 7 cosmological
parameters that characterize a very broad class of CDM models by
statistically confronting the theoretical ${\cal C}_\ell$'s with the
B98 and DMR data.  Sample CDM models that fit the
data are shown in Figure~\ref{powerspectrum}.  These are best-fit
theoretical models using successively more restrictive ``prior
probabilities'' on the parameters. 
A major theme of this paper is to illustrate explicitly how inferences
that are drawn from the CMB data depend on the priors that are
assumed.  Some of
these priors are quite weakly restrictive and are generally 
agreed upon by most cosmologists, for example that the 
Universe is older than 10 Gyr
and that the Hubble constant $H_0 = 100 \, h \mbox{\, km
s}^{-1}\mbox{Mpc}^{-1}$ lies between 45 and 90. 
More strongly restrictive priors rely
on specific measurements, \eg the HST key project determination of
$H_0$ to 10\% accuracy~\cite{hstkeyproj} and the determination of the
cosmological baryon density, $\omega_b\equiv \Omega_b h^2$, to
10\%~\cite{BBN}.  

In \cite{debernardis00}, we applied a ``medium'' set of priors to the B98
power spectrum to constrain a 6 cosmological parameter model and found
a 95\% confidence limit 
for $\Omega_{\rm {tot}}$ of $0.88<\Omega_{\rm {tot}}<1.12$. 
Row P0 of Table~\ref{parameters} shows the result for our full 7 parameter
set with a similar medium prior (here taken to be $h = 0.65\pm 0.1$,
$\omega_b=0.019 \pm 0.006$, with Gaussian errors for both). As we
progress through the Table, we show the effect of either weakening or
strengthening the prior from this starting point.

Two of our parameters are fundamental for describing the physics of
the radiative transport of the CMB through the epoch at $z\sim1100$,
when the photons decoupled from the baryons. These are $\omega_b$ and
the CDM density $\omega_c\equiv \Omega_c h^2$.  The acoustic
patterns at decoupling are related to the sound-crossing distance at
that time, $r_{s}$, which is sensitive to these parameters.  We fix
the density of photons and neutrinos~\cite{mnu}, which are other
important constituents at this epoch.
The observed B98 patterns are also sensitive to the ``angular diameter
distance'' to photon decoupling, mapping the $z\sim1100$ 
spatial structure to the angular structure, and, through its
dependence on geometry, to $\Omega_{\rm {tot}}$, the total energy in units
of the critical density. 
When $\Omega_{\rm {tot}}<1$ (open models), $r_{s}$
is mapped to a small angular scale; when $\Omega_{\rm {tot}}>1$ (closed
models), $r_{s}$ is mapped to a large angular scale. 

This mapping also depends upon the density associated with a
cosmological constant, $\Omega_\Lambda$, and $\Omega_m \equiv
(\omega_c+\omega_b)/h^2$, as well as on 
$\Omega_k\equiv 1 -\Omega_{\rm {tot}}$, the density 
associated with curvature. Combinations of
$\Omega_{k}/\Omega_{m}$ and $\Omega_\Lambda/\Omega_{m}$ which give the
same ratio of angular diameter distance to sound horizon will give
nearly identical CMB patterns, resulting in a near degeneracy that is
broken only at large angular scales where photon transport through
time-varying gravitational potentials plays a role.  One implication
of this is that $\Omega_\Lambda$ cannot be well determined by our data
alone, in spite of the high precision of B98.  We have paid special
attention to such near-degeneracies~\cite{EB98} throughout our
analysis.

The universe reionized sometime between photon decoupling and $z\sim
5$. This suppresses ${\cal C}_\ell$ at small scales by a factor
$e^{-2\tau_c}$, where $\tau_c$, our fifth parameter, is the 
optical depth to Thompson scattering 
from the epoch at which the universe reionized to the present.

Our last two parameters characterize the nature of the fluctuations
arising in the very early universe, through a power law ``tilt'' $n_s$ and
an overall amplitude factor for the primordial perturbations. The
simplest inflation models have a nearly scale invariant spectrum
characterized by $n_s \approx 1$. Of course, many more variables, and
even functions, may be needed to specify the primordial fluctuations,
in particular those describing the possible contribution of gravity
waves, whose role we have also tested~\cite{gw}.  For our overall
amplitude parameter, we use $\ln {\cal C}_{10}$ where ${\cal
C}_{10}$ is the CMB power in the
theoretical spectrum at $\ell = 10$. If we wish to relate the CMB
data to large scale structure (LSS) observations of the Universe, we
use $\ln \sigma_8^2$ as the amplitude parameter, where $\sigma_8^2$ is
the model power in the density fluctuations on the scale of clusters of
galaxies ($8 h^{-1}\mbox{Mpc}$).

Our adopted parameter space is therefore
$\{\omega_b,\omega_c,\Omega_{\rm {tot}},\Omega_\Lambda,n_s,\tau_c,\ln{\cal
C}_{10}\}$. The amplitude ${\cal C}_{10}$ is a continuous variable, and
the rest are discretized for the purpose of constructing the model 
database we use to compare data and theory. 
The number of values and coverage are:
15, over $0.1 \le \Omega_{\rm {tot}} \le 1.5$; 14, over $0.0031\le \omega_{b} \le
0.2$; 10, over $0.03 \le \omega_c  \le 0.8$; 11, over $0 \le
\Omega_\Lambda \le 1.1$; 9, over $0 \le \tau_c \le 0.5$; 31, over $0.5
\le n_s \le 1.5$.  The spacings in each dimension are uneven,
designed to concentrate coverage in the regions preferred by
the data and yet still map the outlying regions\cite{database}. 
Fast CMB transport programs~\cite{CMBFASTCAMB} were used
to construct our ${\cal C}_\ell$ databases. Use was made of
the angular-diameter distance degeneracy and $\ell$-space
compression to reduce the size and computational requirements needed
to construct such a database.  

Parameter estimation is an integral part of the B98 analysis
pipeline, which makes statistically
well-defined maps and corresponding noise matrices
from the time-ordered data, from which we
compute a set of maximum likelihood bandpowers, ${\cal C}_B$.  The
likelihood curvature matrix ${\cal F}_{BB^\prime}$ is calculated 
to provide error estimates including correlations between bandpowers.
The curvature matrix ${\cal F}_{BB^\prime}$ and the curvature
matrix evaluated at zero signal, ${\cal F}_{BB^\prime}^{0}$, are 
used in the offset-lognormal approximation~\cite{BJK98} to compute
likelihood functions $L(x,\vec{y}) = P(\vec{\cal C}_B|x,\vec{y})$
for each combination of
parameters $x$ and $\vec{y}$ in our database.
Here $x$ is the value of the
parameter we are limiting, $\vec{y}$ specifies the values of the other
parameters. 

We multiply the likelihood by our chosen priors,
and marginalize over the values of the other parameters $\vec{y}$,
including the systematic uncertainties in the beamwidth and
calibration of the measurement~\cite{beam}.  This yields the marginalized
likelihood distribution 
\begin{equation}
\mathcal{L}(x) \equiv
        P(x|\vec{\cal C}_B) =
        \int P_{prior}(x,\vec{y}) L(x,\vec{y})  d\vec{y}.
\label{likelihood1}
\end{equation}
For clear detections, central values and $1\sigma$ limits for the
explicit database parameters mentioned above are found from the 16\%,
50\% and 84\% integrals of $\mathcal{L}(x)$.  When no clear detection
exists, these errors can be misleading, so for these cases we shift to
likelihood falloffs by $e^{-1/2}$ from the maximum, or variances about
the mean of the distribution $\mathcal{L}(x)$.  The mean and variance
are used to set the limits on other ``auxiliary'' parameters such as
$h$ and $\Omega_m$, which may be nonlinear combinations of the
database variables.  For good detections the three estimation methods
give very good agreement, and yield 2$\sigma$ errors that are roughly
twice the 1$\sigma$ ones generally reported in this paper.

\begin{figure*}
 \rotatebox{-90}{
  \resizebox{!}{7.8in}{
  \includegraphics{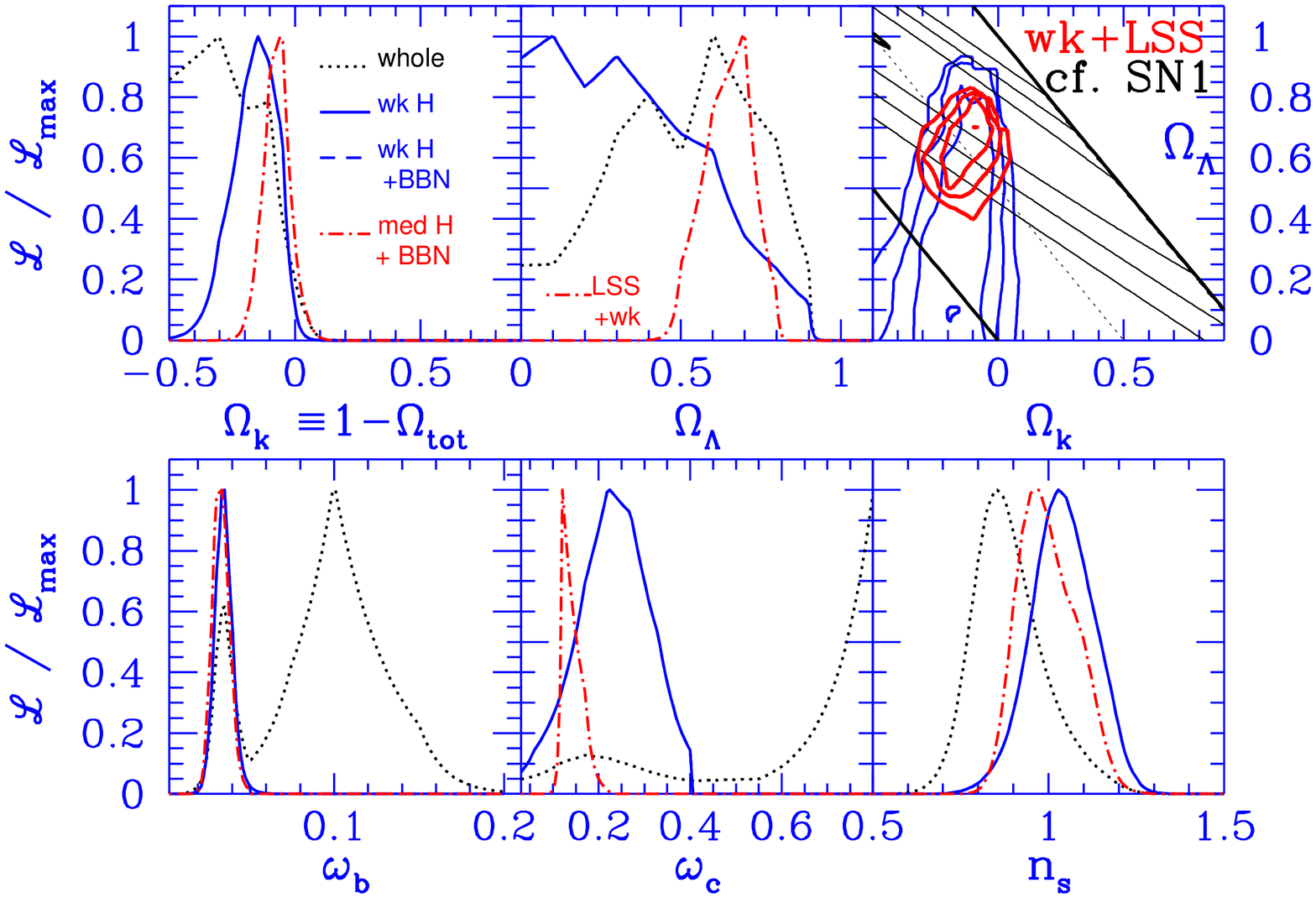}
  }
 }
\caption{\protect\small 
Likelihood functions for a subset of the priors used in
Table~\ref{parameters}. Panel 1 (top left) shows the likelihood for $\Omega_k
\equiv 1-\Omega_{\rm {tot}}$; the full-database (P1, dotted line)
prefers closed models, but reasonable priors (P2, dashed blue line;
P4, solid blue line; P0, dot-dash red line; note that P2 and P4
lie on top of one another in every panel in 
this plot but are distinct in Figure~\ref{6-panel-prior})
progressively move toward $\Omega_k =0$. 
We caution the reader
against agressively interpreting any 2$\sigma$ effects.  Likelihood
curves for $\Omega_{\Lambda}$ are shown in panel 2 (top center). In
panels 2 and 4-6, the cases and line types are as in panel 1, except
that dot-dashed now denotes the weak+LSS prior, P5.  With weak priors
applied, there is no significant detection of $\Omega_{\Lambda}$ (P2
and P4, overlapping as solid blue line in all 
remaining $\mathcal{L}(x)$ panels).  
Only by adding the LSS prior is $\Omega_{\Lambda}$ localized away from zero
(P5, red dot-dash in all remaining $\mathcal{L}(x)$ panels).  Panel 3 (top
right) shows the contour plot of $\Omega_k$ and $\Omega_{\Lambda}$,
for which the first two panels are projections to one axis.  The bold
diagonal black lines mark $\Omega_m$=1 and $\Omega_m$=0.  The blue
contours are those found with the weak prior (P4), plotted at 1, 2, and
3$\sigma$\cite{contourerrors}.  
Red contours are similarly plotted for the weak+LSS
prior (P5).  SN1a constraints are plotted as the lighter
(black) smooth contours, and are consistent with the CMB contours at
the $1\sigma$ level.  
Panel 4 (bottom left) shows the contours
for $\omega_b$; the full database analysis results in a bimodal
distribution with the higher peak concentrated at very high values.
These high $\omega_b$ models are eliminated by the application of a
weak $h$ prior or weak BBN prior (P2 and P4, overlapping as blue here).  
Panel 5 (bottom center) shows a
localization of $\omega_c$ for the weak $h$ and BBN prior cases, but
this is partially due to the effect of the database structure coupling
to the $h$ and age priors.  Only the LSS prior (P5, red dot-dash) allows
the CMB to significantly constrain $\omega_c$.  Panel 6 (bottom right)
shows good localization and consistency in the $n_s$ determination
once any priors are applied.  The inflation-motivated $\Omega_{\rm {tot}}$=1
priors (P10, P11) give very similar curves localized around unity.
See Figure~\ref{6-panel-prior} to see the effects of the database
and priors on these curves.
\label{6-panel} } 
\end{figure*}

We have used this method to estimate parameters, using the B98 power
spectrum of Figure~\ref{powerspectrum} with the COBE bandpowers
determined by \cite{BJK98} and a variety of priors.  The results are
summarized in Table~\ref{parameters};  likelihood functions for
selected parameters and priors are shown in Figure~\ref{6-panel}.

In the presence of degenerate and ill-constrained
combinations of parameters, as with CMB data, the edges of the
data-base form implicit priors.  We have constructed our database such
that these effective priors are extremely broad.  This allows us to
probe the dependence of our results on individually imposed priors.
The choice of measure on the space is itself a prior; 
we have used a linear measure in each of our
variables~\cite{measure}.  
Sufficiently restrictive priors can break parameter degeneracies and
result in more stringent limits on the cosmological parameters.
Artificially restrictive databases or priors can lead to misleading
results; thus, priors should be both well motivated and tested for
stability.  
We therefore regard it as essential that the role of
``hidden priors'' in any choice for ${\cal C}_\ell$ database
construction be clearly articulated. 

\begin{figure*}
 \rotatebox{-90}{
  \resizebox{!}{7.8in}{
  \includegraphics{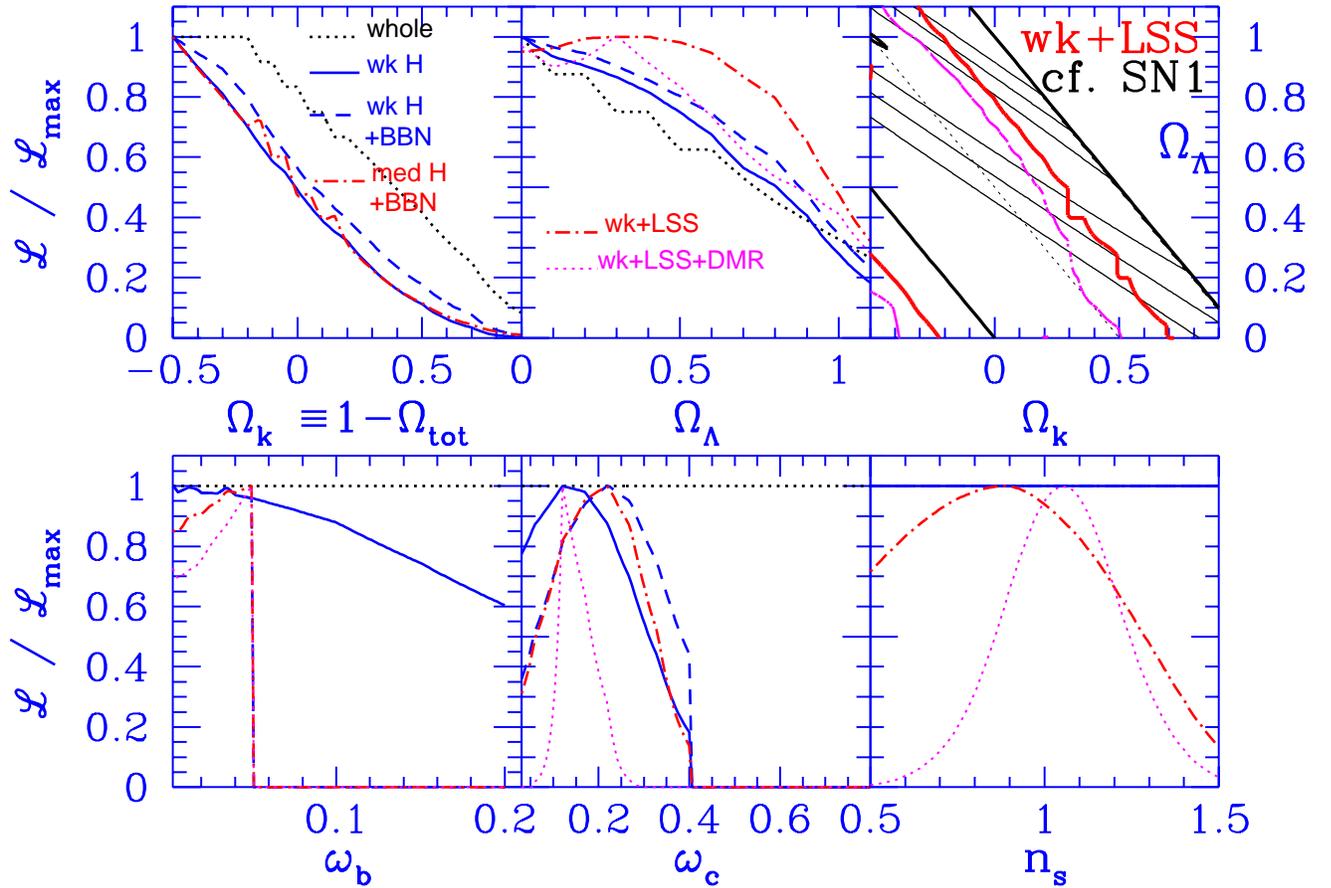}
  }
 }
\caption{\protect\small
Likelihood functions similar to those in Figure~\ref{6-panel}, but
computed without using the B98 data.  These curves show the effect of
the database constraints and applied priors alone.  The identification
of the curves is the same as in Figure~\ref{6-panel}, with the
addition of the dotted magenta curve in panels 2-6, which shows the
likelihood given weak priors and the COBE DMR data. In panel 3, only
the 1$\sigma$ (red) contour is shown for the prior only and prior+DMR cases,
while 1, 2 and 3$\sigma$ (light black) contours are shown for SN1a.
The curves for P2 (solid blue) and P4 (dashed blue) are slightly
separated in this figure, in contrast to Figure~\ref{6-panel}, where they
overlapped.  Of particular interest here are the slope induced across
$\Omega_k$, the slight localization of $\Omega_c h^2$ with the weak
priors, and the significant localization of $\Omega_c h^2$
and $n_s$ with just weak+DMR+LSS (dotted magenta).
\label{6-panel-prior} } 
\vspace{4cm}
\end{figure*}

To illustrate the effects of the database structure and
applied priors, we have plotted likelihood functions found 
using only the database and priors (and no B98 data) in 
Figure~\ref{6-panel-prior}.  These should be compared with
those plotted in Figure~\ref{6-panel} which include the B98
constraints, as discussed below.  We now turn to the 
results found by applying different priors, in the general order of
weakest to strongest applied priors. 

Our ``entire database" analysis prefers closed models with very high
$\omega_b$, as shown in line P1 of Table~\ref{parameters} and in
Figure~\ref{6-panel}.  The low sound speed of these models couples
with the closed geometry to fit the peak near $\ell \sim 200$.  These
models require very high values of $h$ and $\omega_b$, and have
extremely low ages, so we have mapped out this region using a coarse
grid.  The dual-peaked projected likelihood functions shown are
reflections of the the complexity of the full 9-dimensional likelihood
hypersurface.  We note that parameter combinations that 
appear to have a low probability based on 
the projected one-dimensional limits can fit the data quite well,
{\it e.g.}, the $\Omega_{\rm {tot}}=1$ best-fit model of
Figure~\ref{powerspectrum}. 

Applying weakly restrictive priors 
(lines P2-P4 in Table~\ref{parameters}) moves the
result away from the low sound speed models, to a regime which is
stable upon application of more restrictive priors,
as shown in panels 1 (top left) and 4 (bottom left) of
Figure~\ref{6-panel}.  Given their gross conflict with many other
cosmological tests we do not advocate the ``entire database'' models
as representative of the actual universe, and we proceed with
prior-limited analyses below.

The analysis with weakly restrictive priors (P2-P4) finds that the curvature is
consistent with flat, while favoring slightly closed models.  
The migration toward $\Omega_k = 0$ as more restrictive
priors are applied, as shown in Table~\ref{parameters} and in 
panel 1 (top left) of Figure~\ref{6-panel}, suggests caution in 
drawing any conclusion from the magnitude of 
the likelihood drop at $\Omega_k = 0$.  In fact, as is evident from 
Figure~\ref{powerspectrum}, there are models with 
$\Omega_k = 0$ that fit the data quite well.  The likelihood curve 
simply indicates that there are more models with $\Omega_k < 0$ than 
with $\Omega_k > 0 $ that fit the data well.

  We  have taken special care to study the effect on the
likelihood distributions of choosing a different parameterization of
our database.   
For example, we have investigated the likelihoods 
that result from a finely gridded database that uses 
{$\Omega_c$, $\Omega_b$, and $h$} in place
of {$\Omega_k$, $\omega_b$, and $\omega_c$} to determine $\Omega_k$.
This second method, restricted
to $\tau_c =0$ models, uses these different variable choices, 
gridding, and a completely different procedure and code
which uses maximization of the likelihood over other
variables rather than marginalization. 
To compare with this second method, we have taken the 
database used for the table and mimicked the effective priors
due to the parameter limits of the second database.
The results found using these 
two parameterizations and codes agree quite well - 
in all cases the likelihood 
curves shift by at most a small fraction of their width.  
For example, for the applied priors of P2 
the 95\% confidence limits on $\Omega_{{\rm tot}}$ shift from 
$0.99 < \Omega_{{\rm tot}}< 1.24$
for the method used in the table to 
$0.94 < \Omega_{{\rm tot}}< 1.27$
for the {$\Omega_c$, $\Omega_b$, and $h$} method.
Due to the very steep slope of the likelihood near 
$\Omega_k = 0$, however, even this small shift changes 
$L(\Omega_k = 0)/L_{max}$ from 0.2 to 0.8.  
We also find $L(\Omega_k = 0)/L_{max} \approx 0.8$
if we use maximization, rather than
marginalization, in the code used for the table.
Additionally, we note that a downward shift of 
about 5\% in $\Omega_{{\rm tot}}$
occurs if the 10 Gyr age constraint is removed from P2.
These points, plus the
obvious compatibility of the data with the best-fit $\Omega_{tot}=1$
curve in Fig.~\ref{powerspectrum}, reinforces our conclusion 
that there is no significant evidence in the B98 data for 
non-zero curvature.  The 
only valid conclusion to draw from the data that we analyze in this 
paper is that the geometry of the universe is very close to flat.

The baryon density $\omega_b$ is also well constrained.  
While our results are higher ($\sim 3 \sigma$)
than the $\omega_b$ estimates
from light element abundances~\cite{BBN}, it is most
remarkable that our entirely independent method yields a result that
is so close to the BBN value.  
The scalar spectral index $n_s$ is very stable once weak
priors are applied, and is near the value expected from inflation.
This weak prior analysis does not yield a significant detection of
$\Omega_{\Lambda}$; the $\Omega_c h^2$ results in
Table~\protect{\ref{parameters}} are suggestive of a detection, but
are at least in part driven by the weak priors acting on limits of the
database~\cite{hiddenpriors1,hiddenpriors2} as shown in 
Figure~\ref{6-panel-prior}. The values 
of $\tau_c$ are in the range
of expectation of the models, but there is no clear detection.

Note that the the weak priors adopt the conservative restriction that the age
of the universe exceeds 10 Gyr.  Without this, the weak $h$ prior 
still allows a contribution, albeit reduced, from the 
high $\omega_b$, low sound speed, low age solution.  
With the age restriction, this 
solution is eliminated, and the weak BBN 
prior ($\omega_b \le 0.5$) does not significantly change the
constraint: thus, the ``weak $h$+BBN+age" (P4) and ``weak $h$+age" 
(P2) rows are essentially identical. 

In row P4a, we add a ``CMB prior'', which is a full
likelihood analysis of all prior CMB experiments combined with B98 and
DMR, including appropriate filter functions, calibration
uncertainties, correlations, and noise estimates for use in the
offset-lognormal approximation~\cite{BJK98}. As would be expected given
the errors we compute on the compressed bandpowers of these
experiments in Figure~\protect{\ref{powerspectrum}} {\it cf.} those
for B98, this CMB prior only slightly modifies the B98-derived
parameters, with $n_s$ the most notable migration.  None the less, as
much previous analysis of the prior heterogeneous CMB datasets has
shown~\cite{otherparamest}, reasonably strong cosmological conclusions
could already be drawn on $n_s$ and $\Omega_{\rm {tot}}$. Row P4b shows
results excluding B98, for the weak prior case, through our
machinery. Though $n_s$ and $\Omega_k$ have detections consistent with
the B98 results, no conclusions can be drawn on $\omega_b$ (though the
``whole database" analysis does pick 
up the high $\omega_b$, $\Omega_{\rm {tot}}$
region).  We note that if $\tau_c \approx 0 $ is enforced, most
variables remain unmoved, but $n_s$, which is well-correlated with
$\tau_c$, moves closer to unity: for P4, P4a, P4b, we would have $n_s =
0.97,1.03,1.02$, respectively, and for P5, P5a, P5b, we would have 
$n_s =0.93, 0.98,0.98$.  A prior probability of $\tau_c$ based on ideas of
early star formation would help to decrease the $n_s$ degree of
freedom.  

The $\Omega_{\rm {tot}}$, $\omega_b$, and $n_s$ results are stable to the
addition of a prior which imposes two constraints derived from 
LSS observations~\cite{BJ98}. The first is an
estimate of $\sigma_8^2$ that requires the theory in question to
reproduce the local abundance of clusters of galaxies. The second is
an estimate of a shape parameter for the density power spectrum
derived from observations of galaxy clustering~\cite{LSS}.  Adding LSS
to the weak $h$ and BBN priors (P5, and panels 2 (top center) and 3
(top right) of Figure~\ref{6-panel}) breaks a degeneracy, yielding a
detection of $\Omega_\Lambda$ that is consistent with
``cosmic concordance" models.  This also occurs when LSS is added to
only the prior CMB data (P5b and \cite{BJ98}). The LSS prior also
strengthens the statistical significance of the determination of
$\Omega_c h^2$.  Panel 3 (top right) 
of Figure~\ref{6-panel} shows likelihood
contours in the $\Omega_k \equiv 1-\Omega_{\rm {tot}}$ vs. $\Omega_\Lambda$
plane.  Here we have plotted the LSS prior (P5), which strongly
localizes the contours~\cite{contourerrors} away from the
$\Omega_\Lambda = 0$ axis, toward a region that is highly consistent
with the SN1a results~\cite{SN1a}.  Indeed, treating the SN1a likelihood
as a prior does not change the results very much, as indicated by 
row P12 and P13 of the Table, to be compared with rows
P5 and P11, respectively.

The use of a strong $h$ prior alone yields results very similar to
those for the weak $h$ case.  The strong BBN prior, however, shifts
many of the results from the weak BBN case.  Our data indicate a
higher $\Omega_b h^2$ than BBN, and constraining it with the BBN prior
shifts the values of several parameters, including $\Omega_c h^2$,
$\Omega_{\Lambda}$, $n_s$, and $\Omega_m$.  Additional ``strong prior"
results (P8-P11) are shown in Table~\ref{parameters}, as an exercise
in the power of combining other constraints with CMB data of this
quality.

A number of the cosmological parameters are highly correlated,
reflecting weak degeneracies in the broad but 
restricted $\ell$-space range that the B98+DMR data covers~\cite{EB98}.  
Some of these degeneracies can be
broken with data at higher $\ell$, as is visually evident in the
radically different behavior of the models of
Figure~\ref{powerspectrum} beyond $\ell \sim 600$.  To understand the
degeneracies within the context of this data, 
we have explored the structure of the parameter 
covariance matrix $\langle \Delta y_i \Delta y_j
\rangle$, both for the database parameters and the ones derived from
them.  They add motivation for the specific parameter choices we have
made~\cite{rcorr}.  Parameter eigenmodes~\cite{BET97,EB98} of the
covariance matrix, found by rotating into principal components,
explicitly show the combinations of physical database variables that
give orthogonal error bars.  A by-product is a rank-ordered set of
eigenvalues, which show that for the current B98 data, 3 combinations
of the 7 parameters are determined to better than
$10\%$~\cite{parameigen}. 

  We conclude that the B98 data are consistent with the
predictions of the basic inflationary paradigm: that the curvature
of the universe is near zero ($\Omega_k=0$) and that the primordial power
spectrum is nearly scale-invariant ($n_s=1$).  The slight preference
that the current data show for closed, rather than open, models is
not, we believe, a statistically significant indication of non-zero
curvature.  A more conclusive statement awaits further analysis of
B98 data, which will increase the precision of the measured power
spectrum, and/or the results from other experiments.

  We measure a strong detection of the baryon density $\Omega_b h^2$, 
a first for determinations of this parameter from CMB
data.   The value that we measure is robust to the choice of prior,
and is both remarkably close to and significantly higher than that
given by the observed light element
abundances combined with BBN theory.
Assuming that $\Omega_{{\rm tot}} = 1$, 
we find $\Omega_b h^2 = 0.031 \pm 0.004$. 

  Finally, we find that combining the B98 data with 
our relatively weak prior representing LSS
observations and with our other weak priors on the Hubble
constant and the age of the universe yields a clear
detection of both 
non-baryonic matter ($\Omega_c h^2 = 0.014^{+ 0.003}_{-0.002}$) 
and dark energy  ($\Omega_\Lambda = 0.66^{+ 0.07}_{-0.09}$)
contributions to the total energy density in the universe.  The
amount of dark energy that we measure is robust to the inclusion of
a prior on $\Omega_{{\rm tot}}$ (shifting to 
$\Omega_\Lambda = 0.67^{+ 0.04}_{-0.06}$
for $\Omega_{{\rm tot}} = 1$), and to the inclusion of the prior likelihood
given by observations of high-redshift SN1a (shifting to 
$\Omega_\Lambda = 0.69^{+ 0.02}_{-0.03}$
when both the $\Omega_{{\rm tot}} = 1$ and the SN1a priors
are included).  The perfect concordance between the completely
independent detections of $\Omega_\Lambda$ from the CMB+LSS data and from
the SN1a data is powerful support for the notion that the universe is
currently dominated by precisely the amount of dark energy necessary
to provide zero curvature.

        The analysis presented here and in \cite{debernardis00} 
makes use of only a
small fraction of the data obtained during the first Antarctic flight
of \boom .  Work now in progress will increase the precision of
the power spectrum from $\ell  = 50$ to 600, and extend the measurements to
smaller angular scales, allowing yet more precise determinations of
several of the cosmological parameters.


\begin{table*}
\caption{\protect\small 
Results of parameter extraction using successively more restrictive
priors.  The confidence intervals are 1$\sigma$, evaluated using
methods described in the text.  The 2$\sigma$ errors are approximately
double the 1$\sigma$ values quoted in most cases; upper limits
are quoted at 2$\sigma$.  The quoted values
are reported after marginalizing over all other parameters. Note that
these combinations are not, and should not be, the parameters of the
``maximum likelihood" best-fit models of
Figure~\protect\ref{powerspectrum}.  The weak $h$ and BBN ($\Omega_b
h^2$) priors are tophat functions (uniform priors) and both include an
additional age~$> 10$~Gyr prior.  The strong priors are Gaussians with
the stated 1$\sigma$ error, and also have weak constraints
imposed on the other variables.  P0 is the medium $h$ + BBN prior used in
\protect{\cite{debernardis00}} and described in the text. The LSS priors are
combinations of Gaussians and tophats~\protect{\cite{LSS}}.  The SN1
prior (P12, P13) includes LSS as well the SN1a likelihood shown in panel 3
of Fig.~\ref{6-panel}.  P4a and P5a show the small effect of
including prior CMB data in our B98+DMR analysis; these should be
contrasted with P4b and P5b, the case of prior CMB data alone.
Columns 1-5 ($\Omega_{\rm {tot}}$ to $\Omega_{\Lambda}$) are predominantly
driven by the CMB data, except for $\Omega_b h^2$ and $\Omega_b$ when
the strong BBN prior (P7-P9) is applied.  Most of the values in
columns 6-10 ($\tau_c$ to Age) are influenced by the structure of the
parameter space and should not be interpreted as CMB-driven
constraints; exceptions are the $\Omega_{c}h^2$ and $\Omega_\Lambda$
results when the LSS prior is applied. An equivalent table that
includes an inflation-inspired gravity wave induced contribution to
the anisotropy~\protect{\cite{gw}} yields remarkably similar
parameters and errors. }
\label{parameters}

\renewcommand{\arraystretch}{1.1}
\begin{tabular}{lrrrrrrrrrr}
\toprule
Priors
& \multicolumn{1}{c}{$\Omega_{\rm {tot}}$}
& \multicolumn{1}{c}{$\Omega_bh^2$}
& \multicolumn{1}{c}{$n_s$}
& \multicolumn{1}{c}{$\Omega_b$}
& \multicolumn{1}{c}{$\Omega_{\Lambda}$}
& \multicolumn{1}{c}{$\tau_c$}
& \multicolumn{1}{c}{$\Omega_c h^2$}
& \multicolumn{1}{c}{$\Omega_m$}
& \multicolumn{1}{c}{$h$}
& \multicolumn{1}{c}{Age}
\\
P0: Medium $h$+BBN
& $1.07^{0.06}_{0.06}$
& $0.030^{0.004}_{0.004}$
& $1.00^{0.08}_{0.08}$
& $0.08^{0.02}_{0.02}$
& $0.37^{0.23}_{0.23}$
& $0.12^{0.16}_{0.09}$
& $0.25^{0.10}_{0.09}$
& $0.72^{0.23}_{0.23}$
& $0.63^{0.06}_{0.06}$
& $11.9^{1.6}_{1.6}$
\\
\colrule
P1: Whole Database
& $1.31^{...}_{0.16}$
& $0.100^{0.031}_{0.043}$
& $0.88^{0.12}_{0.09}$
& $0.10^{0.05}_{0.05}$
& $0.53^{0.22}_{0.27}$
& $0.22^{0.19}_{0.16}$
& \multicolumn{1}{c}{...}
& $0.81^{0.34}_{0.34}$
& $1.08^{0.39}_{0.39}$
& $7.8^{2.9}_{2.9}$
\\
\colrule
P2: Weak $h$ ($0.45 < h < 0.90$)+age
& $1.15^{0.10}_{0.09}$
& $0.036^{0.006}_{0.005}$
& $1.04^{0.10}_{0.09}$
& $0.11^{0.04}_{0.04}$
& $ < 0.83$               
& $0.21^{0.19}_{0.15}$
& $0.24^{0.08}_{0.09}$
& $0.84^{0.29}_{0.29}$
& $0.58^{0.10}_{0.10}$
& $12.7^{2.1}_{2.1}$
\\
P3: Weak BBN ($\Omega_bh^2 \le 0.05$)+age
& $1.16^{0.10}_{0.10}$
& $0.035^{0.006}_{0.006}$
& $1.03^{0.10}_{0.10}$
& $0.16^{0.09}_{0.09}$
& $ < 0.83$             
& $0.21^{0.19}_{0.15}$
& $0.19^{0.10}_{0.09}$
& $0.92^{0.33}_{0.33}$
& $0.52^{0.14}_{0.14}$
& $14.6^{3.9}_{3.9}$
\\
P4: Weak $h$+BBN+age
& $1.15^{0.10}_{0.09}$
& $0.036^{0.005}_{0.005}$
& $1.04^{0.10}_{0.09}$
& $0.11^{0.04}_{0.04}$
& $ < 0.83$               
& $0.21^{0.19}_{0.15}$
& $0.24^{0.08}_{0.09}$
& $0.84^{0.29}_{0.29}$
& $0.58^{0.10}_{0.10}$
& $12.7^{2.1}_{2.1}$
\\
P4a: Weak and prior CMB
& $1.01^{0.09}_{0.09}$
& $0.031^{0.007}_{0.006}$
& $1.06^{0.10}_{0.09}$
& $0.10^{0.04}_{0.04}$
& $<0.79$                  
& $0.24^{0.19}_{0.17}$
& $0.18^{0.07}_{0.06}$
& $0.64^{0.23}_{0.23}$
& $0.59^{0.11}_{0.11}$
& $13.4^{1.9}_{1.9}$
\\
P4b NO B98: Weak and prior CMB
& $1.03^{0.12}_{0.10}$
& $0.024^{0.017}_{0.018}$
& $1.14^{0.12}_{0.13}$
& $0.08^{0.06}_{0.06}$
& $<0.80$                   
& $0.29^{0.16}_{0.19}$
& $0.21^{0.09}_{0.08}$
& $0.71^{0.28}_{0.28}$
& $0.60^{0.11}_{0.11}$
& $12.9^{2.0}_{2.0}$
\\

\colrule
P5: LSS \& Weak $h$+BBN+age
& $1.12^{0.07}_{0.07}$
& $0.034^{0.006}_{0.005}$
& $0.99^{0.10}_{0.08}$
& $0.10^{0.04}_{0.04}$
& $0.66^{0.07}_{0.09}$
& $0.19^{0.21}_{0.14}$
& $0.14^{0.03}_{0.02}$
& $0.48^{0.13}_{0.13}$
& $0.60^{0.11}_{0.11}$
& $14.5^{1.6}_{1.6}$
\\
P5a: LSS \& Weak and prior CMB
& $1.02^{0.09}_{0.08}$
& $0.030^{0.007}_{0.006}$
& $1.05^{0.10}_{0.08}$
& $0.09^{0.04}_{0.04}$
& $0.47^{0.18}_{0.22}$
& $0.22^{0.19}_{0.16}$
& $0.16^{0.05}_{0.04}$
& $0.57^{0.20}_{0.20}$
& $0.60^{0.12}_{0.12}$
& $13.8^{1.7}_{1.7}$
\\
P5b NO B98: LSS \& Weak and CMB
& $1.00^{0.07}_{0.07}$
& $0.028^{0.015}_{0.015}$
& $1.08^{0.11}_{0.11}$
& $0.08^{0.06}_{0.06}$
& $0.58^{0.13}_{0.17}$
& $0.26^{0.17}_{0.18}$
& $0.14^{0.04}_{0.03}$
& $0.44^{0.15}_{0.15}$
& $0.63^{0.12}_{0.12}$
& $13.8^{1.7}_{1.7}$
\\
P6: Strong $h$ ($h = 0.71 \pm 0.08$)
& $1.09^{0.07}_{0.06}$
& $0.036^{0.005}_{0.005}$
& $1.05^{0.09}_{0.09}$
& $0.08^{0.03}_{0.03}$
& $< 0.82$                
& $0.20^{0.19}_{0.15}$
& $0.26^{0.08}_{0.10}$
& $0.71^{0.27}_{0.27}$
& $0.66^{0.07}_{0.07}$
& $11.6^{1.4}_{1.4}$
\\
P7: Strong BBN ($\Omega_bh^2$=$0.019 \pm 0.002$)
& $1.10^{0.05}_{0.05}$
& $0.021^{0.003}_{0.002}$
& $0.85^{0.08}_{0.07}$
& $0.07^{0.02}_{0.02}$
& $0.79^{0.08}_{0.30}$
& $0.09^{0.12}_{0.07}$
& $0.08^{0.07}_{0.03}$
& $0.38^{0.21}_{0.21}$
& $0.54^{0.10}_{0.10}$
& $17.7^{2.9}_{2.9}$
\\
P8: Strong $h$+BBN
& $1.04^{0.04}_{0.04}$
& $0.021^{0.003}_{0.002}$
& $0.87^{0.07}_{0.07}$
& $0.05^{0.02}_{0.02}$
& $0.75^{0.14}_{0.25}$
& $0.08^{0.12}_{0.06}$
& $0.09^{0.09}_{0.03}$
& $0.28^{0.19}_{0.19}$
& $0.68^{0.09}_{0.09}$
& $15.2^{2.2}_{2.2}$
\\
P9: LSS \& Strong $h$+BBN
& $1.04^{0.05}_{0.04}$
& $0.022^{0.003}_{0.002}$
& $0.92^{0.06}_{0.06}$
& $0.05^{0.02}_{0.02}$
& $0.66^{0.05}_{0.07}$
& $0.08^{0.12}_{0.06}$
& $0.14^{0.03}_{0.02}$
& $0.39^{0.07}_{0.07}$
& $0.64^{0.08}_{0.08}$
& $14.0^{1.3}_{1.3}$
\\
\colrule
P10: $\Omega_{\rm {tot}}=1$ \& Weak $h$+age
& \multicolumn{1}{c}{1}
& $0.031^{0.004}_{0.004}$
& $0.99^{0.07}_{0.07}$
& $0.06^{0.02}_{0.02}$
& $< 0.78$                     
& $0.10^{0.13}_{0.07}$
& $0.27^{0.05}_{0.07}$
& $0.57^{0.21}_{0.21}$
& $0.74^{0.09}_{0.09}$
& $10.9^{0.8}_{0.8}$
\\
P11: $\Omega_{\rm {tot}}=1$ \& LSS \& Weak
& \multicolumn{1}{c}{1}
& $0.030^{0.004}_{0.004}$
& $0.96^{0.07}_{0.06}$
& $0.05^{0.01}_{0.01}$
& $0.67^{0.04}_{0.06}$
& $0.09^{0.12}_{0.07}$
& $0.18^{0.02}_{0.02}$
& $0.32^{0.05}_{0.05}$
& $0.79^{0.05}_{0.05}$
& $11.7^{0.4}_{0.4}$
\\
\colrule
P12: LSS \& Weak \& SN1a
& $1.08^{0.05}_{0.05}$
& $0.034^{0.005}_{0.004}$
& $1.02^{0.09}_{0.08}$
& $0.08^{0.03}_{0.03}$
& $0.72^{0.05}_{0.04}$
& $0.23^{0.19}_{0.17}$
& $0.15^{0.03}_{0.03}$
& $0.37^{0.07}_{0.07}$
& $0.70^{0.09}_{0.09}$
& $13.3^{1.3}_{1.3}$
\\
P13: $\Omega_{\rm {tot}}=1$ \& LSS \& Weak \& SN1a
& \multicolumn{1}{c}{1}
& $0.030^{0.003}_{0.004}$
& $0.97^{0.07}_{0.06}$
& $0.05^{0.01}_{0.01}$
& $0.69^{0.02}_{0.04}$
& $0.10^{0.12}_{0.07}$
& $0.18^{0.02}_{0.01}$
& $0.31^{0.03}_{0.03}$
& $0.81^{0.03}_{0.03}$
& $11.6^{0.3}_{0.3}$
\\
\botrule
\end{tabular}

\end{table*}

{\it Acknowledgements}:

The \boom \ program has been supported 
by NASA (NAG5-4081 \& NAG5-4455),
the NSF Science \& Technology Center for Particle 
Astrophysics (SA1477-22311NM under AST-9120005) and
NSF Office of Polar Programs (OPP-9729121) in the USA,
Programma Nazionale Ricerche in Antartide, Agenzia Spaziale Italiana and 
University of Rome La Sapienza in Italy, and by PPARC in UK.
We thank Saurabh Jha and Peter Garnavich for supplying the SN1a curves
used in Figure~\ref{6-panel}.

\end{document}